\documentclass[11pt]{article}
\usepackage{hyperref,natbib,fullpage,amsmath,amssymb,graphicx}

\newcommand{\etr}{ \ensuremath{ {\rm etr}  }   }

\begin{document}
\title{Simulation of the matrix Bingham-von Mises-Fisher distribution, 
with applications to multivariate and relational data}
\author{Peter D. Hoff \thanks{Departments of Statistics, Biostatistics and the
Center for Statistics and the Social Sciences,
University of Washington,
Seattle, Washington 98195-4322.
Web: \href{http://www.stat.washington.edu/hoff/}{\tt http://www.stat.washington.edu/hoff/}. This work was partially funded by NSF grant number SES-0631531. }
        }
\date{ \today }
\maketitle

\begin{abstract}
Orthonormal matrices play an important role in
reduced-rank matrix approximations and the
analysis of matrix-valued data. 
A matrix Bingham-von Mises-Fisher distribution
is a probability distribution on the set of orthonormal matrices
that includes linear and quadratic terms, and arises
as a posterior distribution in latent factor models
for multivariate and  relational data. 
This article describes rejection and Gibbs sampling 
algorithms for sampling
from this family of distributions, 
 and illustrates their use in the analysis of
a protein-protein interaction network.

\vspace{.2in}
\noindent {\it Some key words}:
Bayesian inference, eigenvalue decomposition,
Markov chain Monte Carlo 
random matrix, social network, Stiefel manifold.
\end{abstract}

\section{Introduction}
Normal vectors and orthonormal matrices play an important role 
in spatial statistics, multivariate analysis and matrix decomposition 
methods.  The set of rank-$R$ $R\times m$ orthonormal matrices is 
called the Stiefel manifold and is denoted $\mathcal V_{R,m}$. 
Probability distributions 
and statistical inference for data from this manifold have 
been developed primarily 
in the spatial statistics literature, particularly for 
the case of points on a sphere ($R=1, m=3$). 
Theoretical treatments of probability distributions on 
higher dimensional manifolds have been given in \cite{gupta_nagar_2000}
and
 \cite{chikuse_2003a}. 

Many of the commonly used probability distributions on $\mathcal V_{R,m}$
have exponential family forms. For example, 
a flexible probability density 
having linear and quadratic terms is given by 
\begin{equation} p_{\rm BMF}(X|A,B,C) \propto \etr( C^TX + B X^T A X), 
\label{eq:1}
\end{equation}
where $A$ and $B$  can be assumed to be symmetric and diagonal matrices,
respectively.
This density, introduced by \cite{khatri_mardia_1977},  
is called the matrix Bingham-von Mises-Fisher density or the 
 matrix Langevin-Bingham density. 

Relevant to the modeling of spatial data, 
\cite{wood_1987} describes a method for simulating from 
densities of this type in cases where $R=1$ and $m=3$, and 
\cite{kent_constable_er_2004} describe methods for  
a complex version of this distribution 
that is feasible  for $R=1$ and small values of $m$. 
 \cite{kume_walker_2006} describe a Gibbs sampling scheme 
for the case $R=1$ and $C=0$. 
However, there is a 
need for statistical and computational tools for data from higher 
dimensional manifolds:
Large, matrix-variate datasets are frequently analyzed and described 
using matrix decomposition techniques, in which heterogeneity across 
rows and columns are represented by low-rank orthonormal eigenvector matrices. 
Probability models for these matrices provide a framework for 
describing variability and uncertainty in matrix variate-data. 
In particular, distributions of the form (\ref{eq:1})
arise as posterior distributions in many models for 
multivariate and relational data:

\paragraph{Example (Factor analysis):}
Let $Y$ be an $n\times p$ data matrix representing $n$
samples from a $p$-variate distribution. If $p$ is large 
or the columns are  highly correlated, it may be desirable to represent $y_i$,  the 
$p$ measurements  within  row $i$, as
linear functions of  $R<p$  latent 
factors $u_i = \{u_{i,1},\ldots, u_{i,R}\}$: 
\begin{eqnarray*} 
y_{i,j}  &=& u_i^T D v_j + \epsilon_{i,j} \\
Y &=& U D V^T +E 
\end{eqnarray*}
In this parameterization, the $n\times R$ and $p\times R$ matrices 
$U$ and $V$ can be assumed to be orthonormal matrices, elements of 
$\mathcal V_{R,n}$ and $\mathcal V_{R,p}$ respectively. 
In situations involving ordinal or missing data  it may be 
desirable to take a likelihood-based approach to estimation
of $U$, $D$ and $V$. 
If the 
error matrix $E$ is made up of independent and identically distributed normal variates, then 
uniform prior distributions  for $U$ and $V$ 
imply that 
\begin{eqnarray*}
p(U | Y,D, V ) &\propto& \etr( [ Y^T VD ]^TU/\sigma^2)  \\
p(V | Y,D, U ) &\propto& \etr( [ Y U D ]^TV/\sigma^2) ,
\end{eqnarray*}
which are matrix von Mises-Fisher distributions. Joint posterior
inference for $U$ and $V$ can be obtained by iteratively  sampling 
from these two distributions. 

\paragraph{Example (Principal components):}
Again, let $Y$ be an $n\times p$ data matrix where the rows are 
assumed to be independent samples from a mean-zero multivariate 
normal population with covariance matrix $\Sigma$. 
Writing $\Sigma$ via its eigenvalue decomposition
$\Sigma= U \Lambda U^T$, the probability density of the data is
\begin{eqnarray*}
 p(Y | \Sigma) &=& (2\pi)^{-np/2} |\Sigma|^{-n/2}
   \etr( - Y\Sigma^{-1}Y^T/2)  \\ 
&=& (2\pi)^{-np/2} \prod_{j=1}^p \lambda_j^{-n/2}
   \etr( - \Lambda^{-1} U^T Y^T Y U/2 ). 
\end{eqnarray*}
A uniform or matrix von Mises-Fisher prior distribution on 
$U\in \mathcal V_{p,p}$
results in a Bingham-von Mises-Fisher posterior distribution. 
This result could be useful if one wanted to use a non-standard
prior distribution for the eigenvalues of $\Sigma$, or if there 
was prior information about the principal components.

\paragraph{Example (Network data):}
Network data consist of binary measurements on pairs of objects or
nodes.
Such data are often represented as a  graph in which
a link between two nodes indicates the presence of a
relationship of some kind. Alternatively,
the data can be represented  with a binary matrix $Y$
so that $y_{i,j}$ is the 0-1 indicator of a link between
nodes $i$ and $j$ (the diagonal of $Y$ is generally undefined).
One approach to modeling such data is to use a latent
factor model with a probit link:
\begin{eqnarray*}
 y_{i,j} &=& \delta_{(c,\infty)}(z_{i,j})  \\
 z_{i,j} &=& u_i^T \Lambda u_j + \epsilon_{i,j}  \\
 Z &=& U \Lambda U^T +E , 
\end{eqnarray*}
where $E$ is modeled as a symmetric matrix of independent 
standard normal noise, 
$\Lambda$ is a diagonal matrix and $U$ is an element of 
$\mathcal V_{R,m}$, with $R$ generally taken to be much smaller than $m$. 
Such a model can be thought of as
a latent eigenvalue decomposition
for the graph $Y$. Given a uniform prior distribution for $U$, 
we have
\begin{eqnarray*}
p(U | Z , \Lambda ) &\propto & \etr(   Z^T U\Lambda U^T/2 ) \\ 
   &=& \etr(  \Lambda U^T Z U/2 ) , 
\end{eqnarray*}
which is a Bingham distribution with parameters $A=Z/2$ and $B=\Lambda$. 

\medskip

Section 2 of this article describes a rejection sampling method
for the matrix von Mises-Fisher  distribution, and  a Gibbs sampling 
algorithm is provided  for cases in which the rejection method 
is infeasible. 
Section 3 
 presents a Gibbs sampling algorithm for generating 
random matrices of arbitrary dimension from the 
Bingham-von Mises Fisher (BMF) distribution. 
Specifically, I show how to construct a Markov chain in $X\in \mathcal V_{R,m}$, samples 
from which converge in distribution to $p_{\rm BMF}$.
Section 4 implements the sampling algorithms in the context of 
a data analysis  
of the interaction network  of 
270 proteins. 
In this example
 the ability to 
sample from the BMF distribution allows for Bayesian inference and 
estimation. A discussion follows in Section  5. 



\section{Sampling from the von Mises-Fisher distribution}
The vector von Mises-Fisher (MF) distribution on the $m$-dimensional 
sphere has a density with respect to the uniform distribution 
given by 
\[ p_{\rm MF}(x|c) = \frac{||c/2||^{m/2-1} }{\Gamma(m/2) I_{m/2-1}(||c||) } 
           \exp\{ c^T x \},  \]
where $I_{\nu}(\cdot)$ is a modified Bessel function of the first kind. 
\cite{wood_1994} provides a sampling scheme for 
this distribution that is relatively straightforward to implement. 
In this section we show how this ability to sample from 
the vector-valued density can be used to sample from 
matrix MF distribution, having density 
\[ p_{\rm MF}(X|C) = [ {_0}F_1(\frac{1}{2} m; \frac{1}{4}D^2) ]^{-1}
    \etr(C^TX) , \]
where  ${_0}F_1(\frac{1}{2} m; \frac{1}{4}D^2)$ is 
a type of hypergeometric function with a matrix argument \citep{herz_1955}. 
We first present a rejection sampling approach
in which a rejection envelope is constructed out of a product of 
vector-valued densities. A second approach is by iterative Gibbs 
sampling of the columns of $X$. 

\subsection{A rejection sampling scheme}
An important reparameterization of the matrix MF density on $\mathcal V_{R,m}$ 
is 
via the singular value decomposition of $C$,  whereby $C= UDV^T$, with $U$ 
and $V$ 
being $m\times R$ and $R\times R$ orthonormal matrices, and 
 $D$ a diagonal matrix with positive entries.
Using these parameters, the density of $X$ can be written as
$p_{\rm MF}(X|U,D,V)\propto \etr( V D U^T X ) =   \etr( D U^T X V )$. 
This density is maximized at $X=UV^T$, which can be interpreted as the
modal orientation of samples from the population. The entries of $D$ can be interpreted as
concentration parameters, describing how close the samples
are to the mode.

As pointed out in \cite{chikuse_2003a}, one way to generate samples
from the matrix MF distribution is with rejection sampling based
on a uniform envelope: Since the density is maximized in $X$ by
$UV^T$, 
the ratio between $p_{\rm MF}$  and the uniform density $g_u$ on $\mathcal V_{R,m}$
is bounded by
\[ \frac{p_{\rm MF}(X)}{g_u(X)} < \frac{\etr( D U^T [U^T V ] V  )}{ {_0}F_1(\frac{1}{2} m; 
  \frac{1}{4}D^2) } = \frac{\etr(D)}{ {_0}F_1(\frac{1}{2} m; 
  \frac{1}{4}D^2) } . \]
If  independent
pairs $X\sim g_u$ and $u\sim {\rm uniform}(0,1)$ are 
 repeatedly sampled  until $u< \etr( C^TX - D)$, then the resulting $X$ has density $p_{\rm MF}$.
As Chikuse points out, such a procedure will be extremely inefficient
for most $C$ of interest, due to the poor approximation of $p_{\rm MF}$
 by $g_u$.

A much better approximation to $p_{\rm MF}$ can be
constructed from a product of vector von Mises-Fisher densities:
If $H$ is an orthogonal matrix, then samples $X$ from ${\rm MF}(H)$
will be such that each column vector $X_{[,r]}$ is close to
the direction of $H_{[,r]}$.
This suggests generating proposal samples $X_{[,r]}$
from a von Mises-Fisher density with concentration vector $H_{[,r]}$,
although constrained to be orthogonal to the other columns of $X$.
With this in mind,
consider the density on $\mathcal V_{R,m}$ corresponding to the
random matrix $X$ sampled as follows:
\begin{enumerate}
\item sample $X_{[,1]} \sim {\rm MF}( H_{[,1]})$.
\item for $r\in \{ 2,\ldots, R\}$:
\begin{enumerate}
\item construct $N_r$, 
    an orthonormal basis for the null space of $X_{[,(1,\ldots r-1)]}$; 
\item sample $z\sim {\rm MF}( N_r^T H_{[,r]})$; 
\item set $X_{[,r]}= N_r z$.
\end{enumerate}
\end{enumerate}
Straightforward calculations show that the
probability density of the matrix $X$ generated as above  can be expressed as
\[ g(X) = \left\{ \prod_{r=1}^R 
   \frac{||N_r^T H_{[,r]}/2||^{(m-r-1)/2} }{\Gamma(\frac{m-r+1}{2}) I_{(m-r-1)/2}(||N_r^T H_{[,r]}||) } \right  \} \etr\{H^TX\}, \]
with respect to the uniform density on $\mathcal V_{R,m}$.
The ratio of the matrix MF density to  $g$ is then
\[ \frac{p_{\rm MF}(X)}{g(X)} = 
 {_0}F_1(\frac{1}{2} m; \frac{1}{4}D^2)^{-1} 
 \left\{ \prod_{r=1}^R  2^{(m-r-1)/2}\Gamma(\frac{m-r+1}{2})\frac{ I_{(m-r-1)/2}(||N_r^T H_{[,r]}||) }{ ||N_r^T H_{[,r]}||^{(m-r-1)/2}} \right \}. \]
Since $I_{k}(x)/x^k$ is an increasing function of $x$, and
  $|| N_r H_{[,r]} ||\leq ||  H_{[,r]}||$, we have
\[ \frac{p_{\rm MF}(X)}{g(X)} <
 {_0}F_1(\frac{1}{2} m; \frac{1}{4}D^2)^{-1} 
 \left\{ \prod_{r=1}^R 2^{(m-r-1)/2}\Gamma(\frac{m-r+1}{2})\frac{ I_{(m-r-1)/2}(||H_{[,r]}||) }{ ||H_{[,r]}||^{(m-r-1)/2}} \right \}  = K(H). \]
For any particular $X$, the ratio of $p_{\rm MF}(X)/g(X)$ to this upper bound is
\[ 
\frac{p_{\rm MF}(X)}{g(X)K(H)} = 
  \prod_{r=2}^R  \frac{ I_{(m-r-1)/2}( || N_r^T H_r ||) }
                      { I_{(m-r-1)/2}( || H_r ||) }
              \left ( 
                   \frac{ ||H_r|| }{ ||N_r^T H_r|| } 
              \right )^{(m-r-1)/2}
       \]
We use this bound  and ratio in the rejection sampler described below.
If $H$ is not orthogonal, then although the bound is sharp the ratio 
may be extremely small:
Consider the case in which $H_{[,1]}$ and $H_{[,2]}$ are both
equal to the vector $a$.  In this case,
samples of $X$ from $p_{\rm MF}$ will have one or the other of its first two
columns close to the direction of  $a$ with equal probability, whereas
samples from $g$ will generally have $X_{[,1]}$ close to the direction of $a$ and $X_{[,2]}$ orthogonal to it. In this case, $||N_2^T H_{[,2]}||$ will
be much smaller than $||H_{[,2]}||$, making the ratio quite small. 
The remedy to this problem is quite simple:
To sample from the matrix MF distribution with a non-orthogonal
concentration matrix $C$ and singular value decomposition
$C=UDV^T$, first sample a matrix $Y \sim {\rm MF}(H)$  with $H$ being
the orthogonal matrix $UD$, using the above described rejection
scheme, then set $X = Y V^T$.
This procedure is summarized as follows:

\begin{enumerate}
\item Obtain the singular value decomposition $UDV^T$ of $C$ and 
  let $H= UD$; 
\item Sample pairs $\{ u, Y\}$ until
      $u< \frac{p_{\rm MF}(Y)}{g(Y)K(H)}$, using the following scheme:
\begin{enumerate}
\item sample $u\sim$uniform(0,1)
\item  sample $Y_{[,1]} \sim {\rm MF}( H_{[,1]})$, and
 for $r\in \{2,\ldots, R\}$ consecutively, 
\begin{enumerate}
\item construct $N_r={\rm Null}(Y_{[,(1,\ldots r-1)]})$
\item sample $z\sim {\rm MF}( N_r^T H_{[,r]})$
\item set $Y_{[,r]}= N_r z$

\end{enumerate}
\end{enumerate}
\item Set $X = YV^T$
\end{enumerate}



To examine the feasibility of this rejection scheme  a small simulation 
study was performed for $m\in \{ 10,20,200\}$ and 
  $R\in \{2,4,6\}$. For each combination of $m$ and $R$, 
three $m\times R$ matrices were constructed, each having $R$ singular 
values all equal to  $m/2,m$ or $2m$. 
One-hundred samples from each MF distribution were generated using the 
above rejection scheme, and the average number of rejected samples 
are recorded in Table \ref{tab:1}. In general, 
the results indicate that the rejection sampler is a feasible method 
for sampling from the MF distribution for a broad range of values of $m,R$ and 
 $D$. 
However, we see that 
the average  number of rejected samples is generally increasing
in the magnitude of the elements of $D$ and the ratio $R/m$.
For large values of this latter ratio, $||N_r^T H_{[,r]}||$ is
typically a
small fraction of
 $||H_{[,r]}||$  and the ratio $p_{\rm MF}(X)/g(X)$ is then rarely
close to the bound.
Similarly, large $D$ leads to large 
differences between  $||N_r^T H_{[,r]}||$ and $||H_{[,r]}||$
for the higher  values of $r$.

\begin{table}
\begin{center}
\begin{tabular}{ c|ccc||c| ccc|| c| ccc| }
   \multicolumn{4}{c}{$m$=10} &  \multicolumn{4}{c}{$m$=20}  & 
    \multicolumn{4}{c}{$m$=200} \\
   $d$ &  $R=2$ & $R=4$ & $R=6$  & 
   $d$ &  $R=2$ & $R=4$ & $R=6$  & 
   $d$ &  $R=2$ & $R=4$ & $R=6$    \\ \hline
 5  & 0.07 & 0.76 & 5.08 &  10 & 0.11 & 0.63 & 3.80 & 100&0.03&0.68&2.28 \\
 10 & 0.09 & 2.40 & 26.52&  20 & 0.24 & 1.78 & 15.25& 200&0.20&1.45&10.04 \\
 20 & 0.38 & 4.01 & 77.74&  40 & 0.30 & 3.65 & 42.70& 400&0.35&3.40 & 36.52 \\
\end{tabular}
\end{center}
\caption{Average number of rejected samples needed to generate  a
 single  
sample from ${\rm MF}(C)$, for various values of $m$, $R$ and 
   singular values $d$. }
\label{tab:1}
\end{table}

\subsection{A Gibbs sampling scheme}

For large values of  $D$ or  $R$ the rejection 
sampler described above may be prohibitively slow. 
As an alternative, a simple Gibbs sampling scheme can be used to 
construct a dependent Markov chain $\{ X^{(1)},X^{(2)},\ldots\}$
which converges in distribution to $p_{\rm MF}$. 
Such a sequence can be constructed by iteratively sampling the columns of $X$ from 
their full conditional distributions. 

The density of $X\sim {\rm MF}(C)$ can be expressed in terms of
a product over the columns of $X$:
\begin{eqnarray*} 
p_{\rm MF}(X|C) &\propto& \etr(C^TX  )  \\
 &\propto& \prod_{r=1}^R \exp ( C_{[,r]}^TX_{[,r]}  )
\end{eqnarray*}
The columns are not statistically independent of course, because
they are orthogonal with probability one.
As such, we can rewrite  $X$ as  $X = \{ X_{[,1]},X_{[,-1]}\} =\{ N z  , X_{[,-1]} \}$,
where $z\in \mathcal S_{m-1}$ and $N$ is an $m\times(m-1)$ orthonormal basis for the
null space of $X_{[,-1]}$. Note that $N^TN=I$ and so  $z=N^TX_{[,1]}$.
Following Chapter 3 of \cite{chikuse_2003a},
the conditional distribution of $z$ given $X_{[,-1]}$ is given by
\begin{eqnarray*} p(z|X_{[,-1]}) &\propto&
    \exp ( C_{[,1]}^T N z  )    =  \exp ( {\tilde c}^Tz  )     
\end{eqnarray*}
which is a vector MF density.  This fact can be exploited to 
implement a  Gibbs sampler that generates a Markov chain in $X$. 
Given a current value $X^{(s)}=X$, the sampler proceeds 
by construction of 
a new value 
$X^{(s+1)}$ as follows:
\begin{itemize}
\item[] Given $X^{(s)}=X$, perform steps 1, 2 and 3  for each $r\in \{1,\ldots, R\} $ in random order:
\begin{enumerate}
\item let $N$ be the null space of $X_{[,-r]}$ and let $z= N^T X_{[,r]}$;
\item sample $z\sim {\rm MF}( N^T C_{[,r]})$ using the sampling scheme of 
      \cite{wood_1994}; 
\item set $X_{[,r]} = N z$.
\end{enumerate}
\item[] Set $X^{(s+1)}=X$. 
\end{itemize}
Iteration of this algorithm generates a reversible aperiodic Markov chain
  $\{ X^{(1)}, X^{(2)}, \ldots \}$. If $m>R$ this Markov chain is
also irreducible, and samples from the chain converge in distribution to
${\rm MF}(C)$. However, if $m=R$ then the chain is reducible.
 This is
because in this case the null space  of $X_{[,-r]}$ is one dimensional and therefore
$z \in \{-1,1\}$, meaning that the samples from the Markov chain
will remain fixed up to column-wise multiplication of $\pm 1$.
The remedy for this situation is not too difficult:
An irreducible Markov chain for the case $m=R$ can be constructed by
sampling two columns of $X$ at a time. Details of such a procedure  are 
given in the context of the  more 
general Bingham-von Mises Fisher distribution in the next section.

Finally, we note that non-orthogonality among the columns of $C$ can add to 
the  autocorrelation in the Gibbs sampler. This 
source of undesirable  autocorrelation can be removed by 
performing the Gibbs sampler 
on $Y\sim {\rm MF}(H)$, where $X=UDV^T=Y V^T$ and $H=UD$, as 
described in the previous subsection.

\section{Sampling from the Bingham-von Mises-Fisher  distribution}
In this section a Markov chain Monte Carlo method for sampling 
from  
$p_{\rm BMF}$ is derived. 
Similar to the approach in Section 2.2, 
the method involves iteratively 
resampling each column of $X$ given the others using full conditional 
distributions, thereby 
generating a Markov chain $\{ X^{(1)}, X^{(2)}, \ldots
  \}$ whose stationary 
distribution converges to  $p_{\rm BMF}$. 
We first present a method for sampling from the vector BMF distribution
on $x\in \mathcal S_m$, and
then show how this can be used iteratively to sample from the 
matrix valued extension.

\subsection{The vector Bingham distribution}
The  
Bingham distribution  on the $m$-dimensional sphere has a density 
with respect to the uniform distribution given by
\[ p_{\rm B}(x|A) \propto  \exp( x^T A x ), \ \ x\in \mathcal S_m .  \]
 Since 
$x^T A x=x^T A^T x = \frac{1}{2}x^T (A+A^T )x$ , $A$ can 
be assumed  to be  symmetric.
Let the eigenvalue decomposition of $A$ be $A=E \Lambda E^T$, 
and let $y= E^Tx$. Since  $E$ is orthonormal the change of variables 
formula gives the density of $y$ as 
\begin{eqnarray*} p(y|E,\Lambda) &=& c(\Lambda) \exp(y^T E^T E\Lambda E^T E y ) \\
& =&  c(\Lambda)\exp( y^T \Lambda y )   \\
& \propto &  \exp(\sum_{i=1}^m \lambda_i y_i^2  ) . 
\end{eqnarray*}
Again, this density is with respect to the uniform density on the 
sphere. 
If we  write $y_m^2= 1-\sum_{i=1}^{m-1} y_i^2$, 
the uniform density in terms of the unconstrained coordinates
  $\{ y_1,\ldots, y_{m-1}\}$ is proportional 
to $|y_m|^{-1}=({1-\sum_{i=1}^{m-1} 
   y_i^2})^{-1/2}$
 and so the above density with respect to  Lebesgue measure on 
  $y_1,\ldots, y_{m-1}$ becomes 
\[ 
 p(y|E,\Lambda) \propto   \exp(\sum_{i=1}^m \lambda_i y_i^2) 
      |y_m|^{-1}, \  \ y_m^2 = 1-\sum_{i=1}^{m-1} y_i^2 \in [0,1]. \]

We now consider  Gibbs sampling of the vector $y$. 
One possibility would be to sample components of $y$ from their 
full conditional distributions. This is the approach taken by 
\cite{kume_walker_2006}. 
While straightforward, such an approach 
can result in a slowly mixing Markov chain because the full conditionals
are so highly constrained. For example, the full conditional 
distribution of $y_1$ given $y_2,\ldots, y_{m-1}$ 
is non-zero only on $y_1^2 < 1- \sum_{2}^{m-1} y_i^2$. 
As an alternative, let 
$\theta = y_1^2$ and $q = \{ y_1^2/(1-y_1^2),\ldots, y_m^2/(1-y_1^2)\}$, so that $\{ y_1^2,\ldots, y_m^2\} = \{ \theta, (1-\theta)q_{-1} \}$.
Sampling a new value of $y_1^2=\theta \in (0,1)$ 
conditional on $q_{-1}$  allows for larger redistributions of 
the ``mass'' of $\{ y_1^2,\ldots, y_m^2\}$  
 while ensuring that $\sum_1^m y_i^2 =1$. 

Keeping in mind that $y_m^2 =1-\sum_1^{m-1} y_i^2$ and 
  $q_{m}=1-\sum_{2}^{m-1} q_i$, 
the joint density of $\{\theta,q_2,\ldots, q_{m-1}\}$ 
can be obtained from that of $\{y_1,\ldots, y_{m-1}\}$ 
as follows:
 \[ |\frac{ d \theta }{d {y_1}}| = 2 |y_1| =  2\theta^{1/2} \ \   ,  \ \
   |\frac{d q_i}{d y_{i}}| = 2\frac{ |y_i|}{1-y_1^2} = 2q_i^{1/2}(1-\theta)^{-1/2}  ,\  i>1  \]
and so 
\begin{eqnarray*}
 p(\theta,q_{-1}) &\propto& \exp( \theta \lambda_1 + (1-\theta)q_{-1}^T \lambda_{-1})  \times
   | \frac{ d \{\theta,q_{-1}\} }{ d \{ y_1,\ldots, y_{m-1}\} } |^{-1}
  [ (1-\theta)q_m]^{-1/2} \\
&=& \exp( \theta \lambda_1 + (1-\theta)q_{-1}^T \lambda_{-1}) 
 \times\theta^{-1/2}(1-\theta)^{(m-3)/2} \prod_{i=2}^{m} q_i^{-1/2} \\
 p(\theta|q_{-1}) &\propto& \exp ( \theta [\lambda_{1} -q_{-1}^T \lambda_{-1}]) 
   \times\theta^{-1/2}(1-\theta)^{(m-3)/2}. 
\end{eqnarray*}
Sampling $\theta\in (0,1)$  can proceed by computing the 
relative probabilities on a grid or with a rejection scheme as described 
at the end of Section 3.2. 
Iteration  of this procedure with $\theta=y_i^2$ for each $i\in \{1,\ldots, m\}$
and a similarly redefined $q$ generates a Markov chain in $\{ y_1^2,\ldots, 
 y_m^2\}$ with a stationary distribution equal to $p(y_1^2,\ldots, y_m^2|E,\Lambda)$.
The signs of the $y_i$'s do not affect the density and can each be randomly
and independently assigned to be positive or negative with equal probability.
The value of $x$ is then obtained from $x=E y$. 
To summarize, Markov chain Monte Carlo  samples from $p_{\rm B}(x|A)$ can be obtained by 
iterating the following algorithm:

\begin{itemize}
\item[] Given $A=E^T \Lambda E$ and a current value of $x^{(s)}=x$, 
\begin{enumerate}
\item compute $y=E^T x$; 
\item perform steps (a)-(d) for each $i \in \{1,\ldots, m\}$ in random order:
\begin{enumerate}
\item let $\{q_1,\ldots, q_{m}\}=  
        \{ y_1^2/(1-y_i^2),\ldots, y_m^2/(1-y_i^2) \}$; 
\item sample $\theta\in (0,1)$ from the density proportional to  
$e^{\theta (\lambda_{i} -q_{-i}^T \lambda_{-i})}
   \times\theta^{-1/2}(1-\theta)^{(m-3)/2}$; 
\item  sample 
   $s_i$ uniformly from $\{-1,+1\}$; 
\item set $y_i=s_i \theta^{1/2}$  and for each $j\neq i$, set $y_j^2=(1-\theta) q_j$ 
leaving the sign unchanged. 
\end{enumerate}
\item set $x=Ey$.
\end{enumerate}
\item[] Set $x^{(s+1)}=x$. 
\end{itemize}

\subsection{The vector Bingham-von Mises-Fisher distribution}
The Bingham-von Mises-Fisher (BMF) density adds a linear term to the quadratic of the Bingham 
density, so that 
$ p(x|A) \propto  \exp( c^T x +x^T A x  )$. A Gibbs sampling algorithm for this distribution can proceed in nearly the same way as for the 
Bingham distribution. For the BMF distribution, the signs of the 
$y_i$'s are not uniformly distributed on $\{-1,+1\}$, and so 
we parameterize $y$ in terms of  $\theta$ and $q$ as above but additionally 
let $s_i = {\rm sign}(y_i)$. For a given value of the vector $s$ the transformation 
between $(\theta,q_{-1})$ and $y$ is one-to-one, and the conditional density 
of $\{\theta,s_1\}$ given $q_{-1}$ and $s_{-1}$ is 
\begin{eqnarray*}  p(\theta,s_1|q_{-1},s_{-1})  &\propto &  \left \{
e^{ \theta (\lambda_{1} -q_{-1}^T \lambda_{-1}) } 
   \times \theta^{-1/2}(1-\theta)^{(m-3)/2} \right \} \\ 
    & & \times
   \exp( \theta^{1/2}s_1d_1 + (1-\theta)^{1/2}( s_{-1}\circ q_{-1}^{1/2}  )^T 
       d_{-1} ), 
\end{eqnarray*}
where $d=E^Tc$. A value of $\{ \theta, s_1\}$ can be sampled from its full conditional distribution by first sampling $\theta\in (0,1)$ from 
$p(\theta|q_{-1},s_{-1})$ and then 
sampling $s_1$  conditional on $\theta$. 
This results in the  following modification steps 2(b) and 2(c) above:
\begin{itemize}
\item[(b)] sample $\theta$ from the density proportional to
     $p(\theta,s_i=-1|q_{-i},s_{-i}) + p(\theta,s_i=+1|q_{-i},s_{-i})$; 
\item[(c)] sample $s_i\in \{-1,+1\} $ with probabilities proportional to 
 $\{ e^{-\theta^{1/2}d_i},  e^{+\theta^{1/2}d_i} \}$. 
\end{itemize}
Again, the simplest way to sample $\theta$ is by approximating its distribution
on a grid of $(0,1)$. Alternatively, it is not too hard to come up with 
efficient rejection sampling schemes for various  values of the parameters. 
The density we need to sample from is of the form
\[ p(\theta ) \propto 
   \theta^{-1/2} (1-\theta)^{k} e^{\theta a + (1-\theta)^{1/2} b }\times
       ( e^{ - \theta^{1/2} c} +  e^{  \theta^{1/2} c} ).  \]
If $k$ is larger than $a,b$ or $c$ then a beta(1/2,$k$) distribution 
works well as a proposal distribution. 
On the other hand, if  $a$ is much larger than the other parameters then 
there will be a local mode close to 1. I have found that 
 a beta$(1/2, 1+ k\wedge [ (k-a)\vee -1/2 ])$ proposal distribution works well 
for a wide range of scenarios where either $a$ or $k$ dominate the density 
(as they do for the  data analysis in Section 4). 
Further  details of a rejection sampling scheme based 
on this proposal distribution for $\theta$ 
are available in the {\sf R}-software 
companion to this article.

\subsection{The matrix Bingham-von Mises-Fisher distribution}
Expressing the density of $X\sim {\rm BMF}(A,B,C)$  in terms of 
a product over the columns of $X$, we have
\begin{eqnarray*} 
p_{\rm BMF}(X|A,B,C) &\propto& \etr(C^TX + B X^T A X )  \\
 &\propto& \prod_{r=1}^R \exp ( C_{[,r]}^TX_{[,r]} + b_{r,r} X_{[,r]}^TAX_{[,r]} )
\end{eqnarray*}
As in Section 2, the 
columns are not statistically independent because 
they are orthogonal with probability one. 
We rewrite  $X$ as  $X = \{ N z  , X_{[,-1]} \}$ as before, 
where $z\in \mathcal S_{m-1}$ and $N$ is an $m\times(m-1)$ orthonormal basis for the 
null space of $X_{[,-1]}$. 
The conditional density of $z$ given $X_{[,-1]}$ is 
\begin{eqnarray*} p(z|X_{[,-1]}) &\propto&
    \exp ( C_{[,1]}^T N z + b_{1,1} z^T N^TA N z )      =
     \exp ( {\tilde c}^Tz + z^T \tilde A z )  
\end{eqnarray*}
which is a vector BMF density. 
A Markov chain in $X$ that converges to ${\rm BMF}(A,B,C)$ can 
therefore be constructed as follows:
\begin{itemize}
\item[] Given $X^{(s)}=X$, perform steps 1-4 for each $r\in \{1,\ldots, R\} $ in random order:
\begin{enumerate}
\item let $N$ be the null space of $X_{[,-r]}$ and let $z= N^T X_{[,r]}$;  
\item compute $\tilde c = N^T C_{[,r]}$ and $\tilde A = b_{r,r} N^T A N$; 
\item update the elements of $z$ using the Gibbs sampler for the  vector BMF$(\tilde A,\tilde c)$ density; 
\item set $X_{[,r]} = N z$. 
\end{enumerate}
\item[] Set $X^{(s+1)} = X$. 
\end{itemize}
Iteration of this algorithm generates a reversible aperiodic Markov chain
$\{ X^{(1)},X^{(2)},\ldots \}$ which is irreducible for $m>R$ 
 and thus converges in distribution to 
${\rm BMF}(A,B,C)$. However, if $m=R$ then the chain is reducible. 
As described in Section 2.2, 
this is 
because in this case the null space  of $X_{[,-r]}$ is one dimensional 
and  the samples from the Markov chain 
will remain fixed up to column-wise multiplication of $\pm 1$. 
The remedy for this situation is to sample multiple columns of $X$ at 
a time. In particular, the full conditional distribution of two columns 
is easy to derive. 
For example, 
the null space $N$ of $X_{[,-(1,2)]}$ is an $m\times 2$ matrix, 
and  so $X_{[,(1,2)]}= N Z$ where $Z$ is a $2\times 2$ orthonormal 
matrix. 
The density of $Z$ given $X_{[,-(1,2)]}$ is 
\[  p(Z)  \propto  
\etr( {\tilde C}^T Z + \tilde B Z^T \tilde A Z ) \]
where $\tilde C =  N^T C_{[,(1,2)]}$, $\tilde B ={\rm diag}(b_{1,1},b_{2,2})$ and 
$\tilde A =  N^T A N$.
Since $Z$ is orthogonal, we can  parameterize it as
\[ Z = \left ( \begin{array}{rr} \cos \phi & s \sin \phi \\
                          \sin \phi & -s \cos \phi \end{array} \right ) \] 
for some $\phi \in (0,2\pi)$ and $s=\pm 1$. 
The second column $Z_{[,2]}$ of $Z$ is a linear function of 
the first column $Z_{[,1]}$, 
and the uniform density on the circle is constant in $\phi$, 
so the joint density of 
 $(\phi, s)$ is simply $p(Z(\phi,s))$. 
Sampling from this distribution can be accomplished by first sampling 
$\phi \in (0,2\pi)$ from a density proportional to 
$p(Z(\phi,-1) )+p(Z(\phi,+1) )$, and 
then sampling $s$ conditional on $\phi$. 
To summarize, the Gibbs sampling scheme  for the case $m=R$ is as follows: 

\begin{itemize}
\item[] Given $X^{(s)}=X$, perform steps 1-5 for each 
pair $(r_1,r_2) \subset \{ 1,\ldots, R\}$ in random order:
\begin{enumerate}
\item let $N$ be the null space of $X_{[,-(r_1,r_2)]}$; 
\item compute $\tilde C =  N^T C_{[,(r_1,r_2)]}$, $\tilde B ={\rm diag}(b_{r_1,r_1},b_{r_2,r_2})$ and
$\tilde A =  N^T A N$; 
\item sample $\phi\in (0,2\pi)$ from the density  proportional to 
     $p(Z(\phi,-1))+p(Z(\phi,+1) )$; 
\item sample $s \in  \{-1,+1\}$  with probabilities 
proportional to 
 $\{p(Z(\phi,-1) )$, $p(Z(\phi,+1) )\}$; 
\item set $Z=Z(\phi,s)$ and $X_{[,(r_1,r_2)] } = N Z$. 
\end{enumerate}
\item[] Set $X^{(s+1)}=X$. 
\end{itemize}



\section{Example: Eigenmodel estimation for network data}

\begin{figure}
\centerline{\includegraphics[height=3.5in]{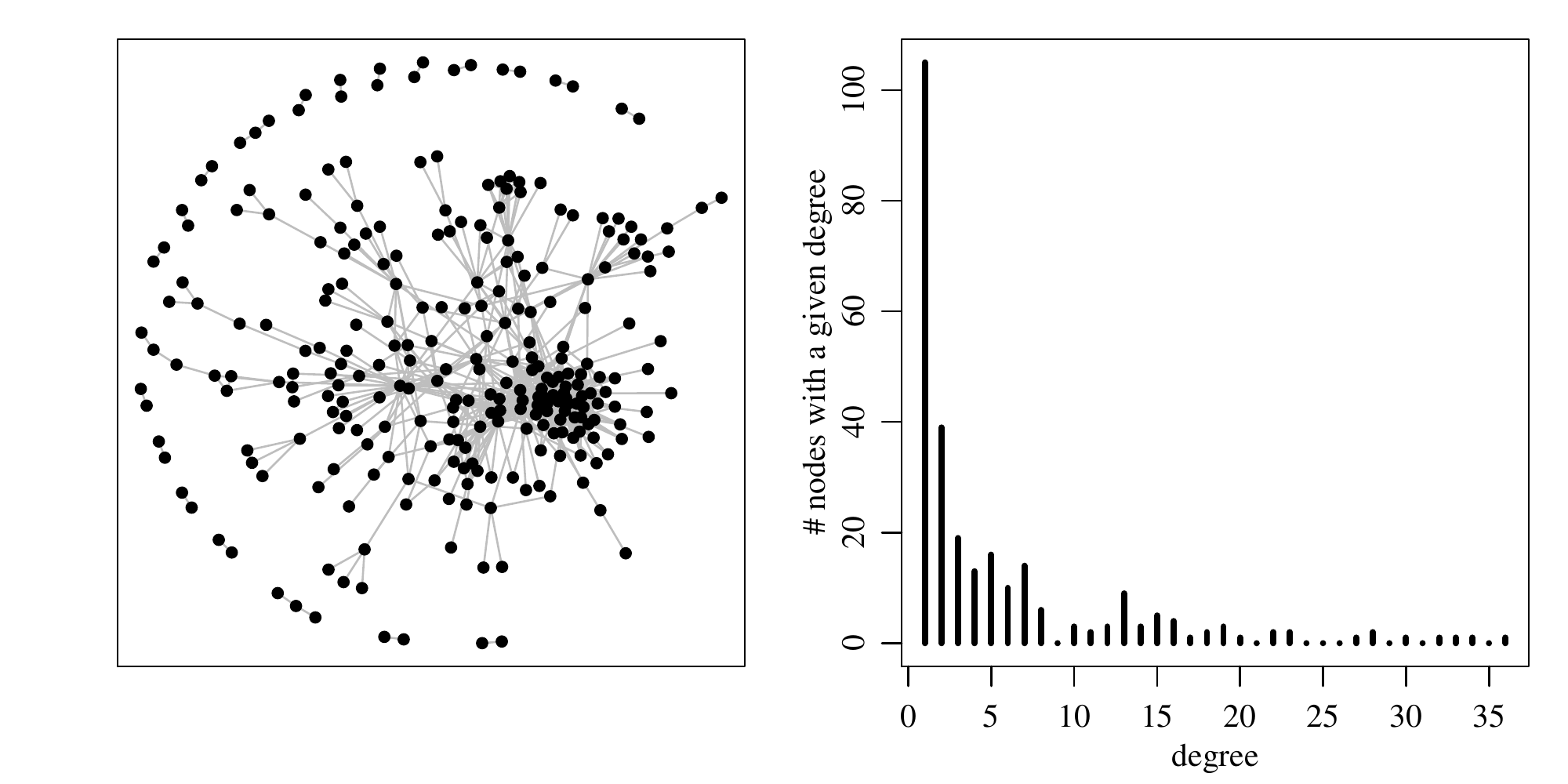}}
\caption{Descriptive plots of the protein interaction network.
The first panel shows the complete dataset. The second panel gives
a histogram of the degree distribution.  }
\label{fig:2}
\end{figure}

In this section we use the model for network data described in 
the Introduction to analyze 
a symmetric binary matrix
of  protein-protein interaction data, originally described in
 \cite{butland_2005}.  For these data,  
 $y_{i,j}= 1 $ if proteins $i$ and $j$ bind together and $y_{i,j}=0$ otherwise.
The data consist of pairwise measurements among $m=270$ essential proteins 
of {\it E.\ Coli}. 
The interaction rate is $\bar y = 0.02$, with  most 
nodes (53\%) having only 1 or 2 links. Nevertheless,
the large connected component of the graph consists of 230 of the 270
nodes, as shown in the first panel of Figure \ref{fig:2}. 

As described in the introduction, 
our model for these data  is 
essentially a latent factor model with a probit link: 
\begin{eqnarray*}
 y_{i,j} &=& \delta_{(c,\infty)}(z_{i,j})  \\
 z_{i,j} &=& u_i^T \Lambda u_j + \epsilon_{i,j}  \\
 Z &=& U \Lambda U^T +E , 
\end{eqnarray*}
With $U \in \mathcal V_{R,m}$, 
this model can be thought of as an  $R$-dimensional  
latent eigenvalue decomposition
for the graph $Y$. 
\cite{hoff_2005} discusses parameter estimation for a version of 
this  model, and \cite{hoff_ward_2004} use such a model to describe 
networks of international relations. However, the approaches in 
these papers use normal distributions for the latent factors, 
leading to some identifiability issues. For example, the magnitudes of the 
factors are confounded with the eigenvalues, and attempts to 
fix the scale of the $u_i$'s lead to complicated constraints among 
their multivariate means, variances and  covariances. 

A cleaner approach to modeling such data is to estimate 
$\{u_1,\ldots, u_m\}$ as being the rows of an $m\times R$ 
orthonormal matrix $U$. The probability density of a symmetric 
matrix $Z$ with mean $U \Lambda U^T$ and off-diagonal unit variance is  
\begin{eqnarray*} p(Z| U,L) &\propto & \etr [-(Z- U \Lambda U^T )^T 
     (Z-U \Lambda U^T)/4  ]\\
 &=& \etr(- Z^TZ/4) \etr(  Z^T U\Lambda U^T/2) 
    \etr( - \Lambda^2/4). 
\end{eqnarray*}
We call this model a latent eigenmodel, as the parameters $U$ and $\Lambda$
are the eigenvectors and values of the mean matrix of $Z$. 
The $-1/4$ has replaced the usual $-1/2$ because $Z$ is symmetric. 
Additionally, the diagonal elements  of $Z$ have variance 2, but do not correspond to any 
observed data as the diagonal of $Y$ is undefined. These diagonal elements 
are  integrated
over 
in the Markov chain Monte Carlo estimation scheme described below.

 Using a uniform prior distribution 
on $U$ and independent normal$(0,\tau^2)$ prior distributions for 
the elements of $\Lambda$ gives
\begin{eqnarray*}
p(\Lambda | Z , U ) &=& \prod_{r=1}^R {\rm dnorm}(\lambda_r:
   {\rm mean}=\tau^2 U_r^TZ U_r/(2 + \tau^2), 
   {\rm var}= 2\tau^2/(2+\tau^2) ) \\
p(U | Z , \Lambda ) &\propto & \etr(   Z^T U\Lambda U^T/2 ) \\ 
   &=& \etr(  \Lambda U^T Z U/2 )  \\
   &=& {\rm dBMF}(U: A=Z/2, B=\Lambda,C= 0)
\end{eqnarray*}
where ``dnorm'' and ``dBMF'' denote the normal and BMF
densities with the corresponding parameters. Approximate 
posterior inference for  $U$ and $\Lambda$ can be obtained via 
iterative Gibbs sampling of $\{ U, \Lambda, Z, c\}$ from their full conditional 
distributions given the data $Y$. One iteration of the sampling 
scheme consists of the following: 
\begin{itemize}
\item[] Given   $\{ U, \Lambda, Z, c\}^{(s)} =\{ U, \Lambda, Z, c\}$
\begin{enumerate}
\item sample the columns of $U$
      from their full conditional distributions under 
      ${\rm BMF}(Z/2, \Lambda,0)$; 
\item sample the elements of $\Lambda $ from their  normal conditional 
   distributions given above; 
\item sample the elements of $Z$ from 
       normal densities with mean $U\Lambda U^T$ but constrained to be above 
       or below $c$ depending on $Y$; 
\item sample $c$ from 
      a constrained normal distribution. 
\end{enumerate} 
\item[] Set  $\{ U, \Lambda, Z, c\}^{(s+1)} =\{ U, \Lambda, Z, c\}$. 
\end{itemize}
A natural choice of the prior parameter $\tau^2$ is $m$, as this 
is roughly the variance of the eigenvalues of an $m\times m$ matrix 
of independent standard normal noise. 

There are several reasons for fitting a statistical model to these data:
First of all, the undefined diagonal $\{y_{i,i} \}$  
precludes a standard eigenvalue 
decomposition of the original data. Second, even if the diagonal could 
be reasonably defined, the data are binary and so a decomposition on 
this raw data scale may be inappropriate. 
Additionally, a statistical model provides 
measures of uncertainty and predictive probabilities. The latter can be 
particularly useful in terms of outlier analysis:  $\{ i,j\}$ pairs for 
which $y_{i,j}=0$ but $\hat \Pr (y_{i,j}=1)$ is large might 
indicate a ``missing link''
and could warrant further investigation.

\begin{figure}
\centerline{\includegraphics[height=3.5in]{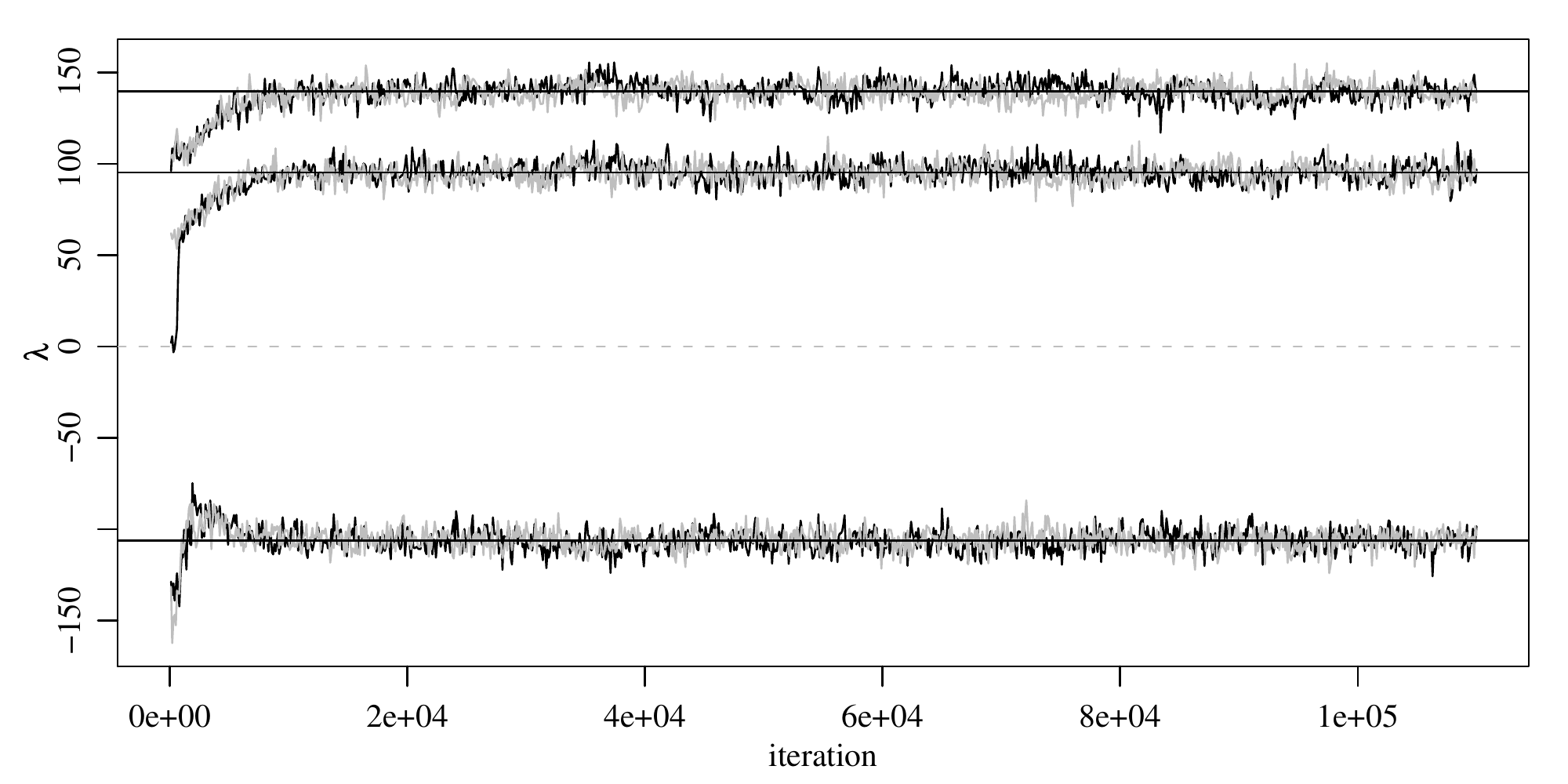}}
\caption{Samples of $\Lambda$ from the two independent Markov chains. }
\label{fig:3}
\end{figure}

A three-dimensional eigenmodel was fit to the protein interaction
 data using the 
Gibbs sampling scheme described above. 
Specifically, two independent Gibbs samplers of length 110,000 were 
constructed, one initiated with random starting values and the 
other with values obtained from the eigenvalue decomposition 
of a rank-based transformation of $Y$, 
in which the ranks of tied values were randomly assigned.
Both chains converged to 
the same region of the parameter space after a few thousand iterations. 
A plot of the sequences of   sampled 
eigenvalues from each of the chains is given in Figure {\ref{fig:3}}, 
indicating one negative and two positive eigenvalues. 
For posterior analysis the first 10,000 iterations of each chain were dropped
and only every 100th iteration was retained, leaving 1000 sample values 
from each of the two chains. 
From these samples we can calculate the posterior mean value of $U\Lambda U^T$, 
which is not exactly a rank 3  matrix 
but is very close - its first three eigenvectors accounted for 
 99.95 percent of the sum of squares of the posterior mean. 
The eigenvectors corresponding to the largest eigenvalues of 
this mean matrix 
 can  be reasonably thought of as a posterior 
point estimate   of $U$. 

\begin{figure}
\centerline{\includegraphics[height=3.5in]{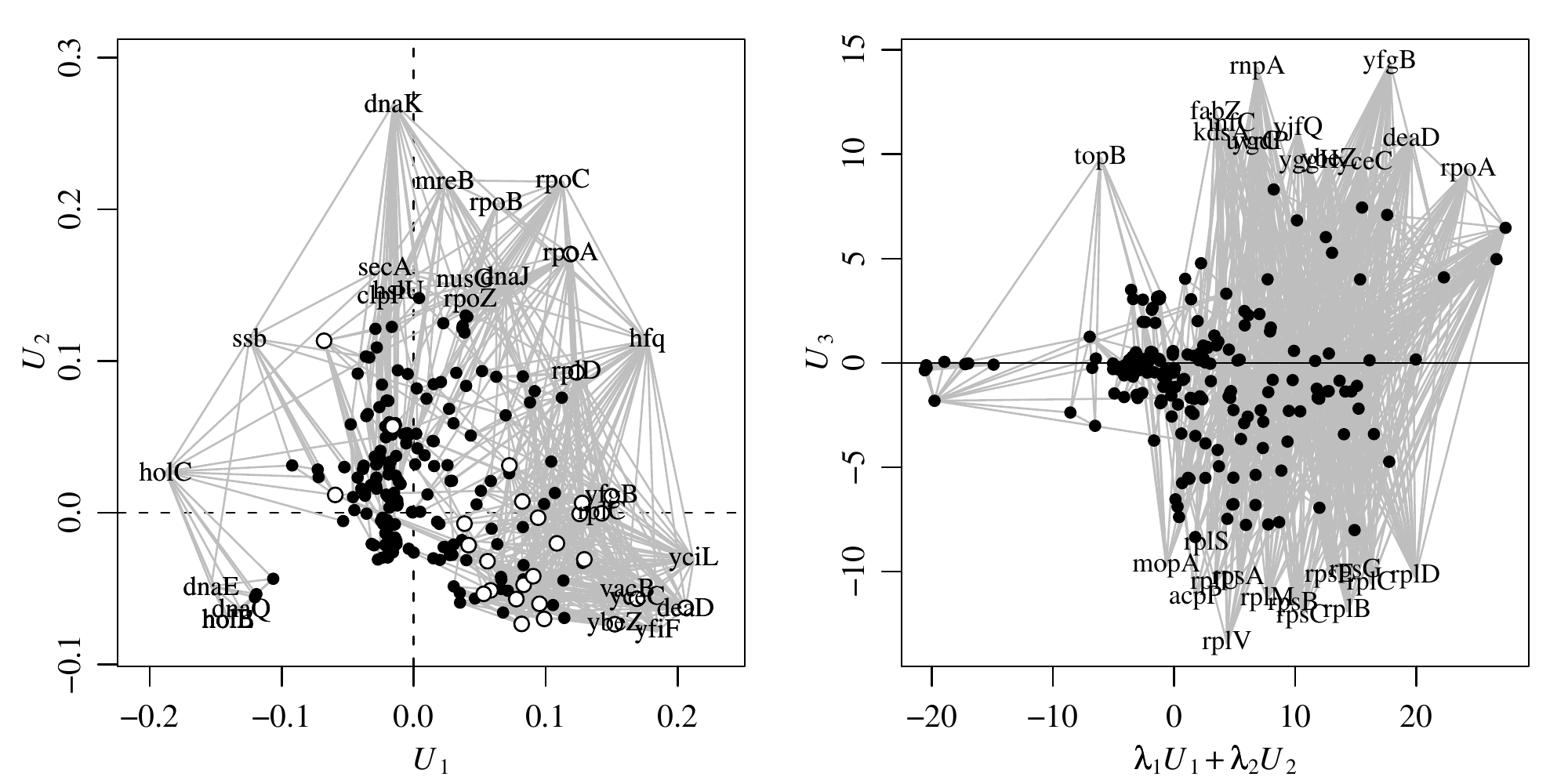}}
\caption{Plots of the latent eigenvectors. In the first panel, 
  nodes with large values of $|u_{i,3}|$ are plotted with white 
circles.    }
\label{fig:4}
\end{figure}

The eigenvectors corresponding to the two positive eigenvalues are 
plotted in the first panel of Figure \ref{fig:4}, along with links 
between interacting protein pairs. Proteins with large values of 
$u_{i,1}^2+u_{i,2}^2$ 
are plotted using their names.
For positive eigenvalues, the interpretation of the parameters is 
that $(u_{i,1},u_{i,2})$ and $(u_{j,1},u_{j,2})$ being in the 
same direction contributes to the tendency for there to be an interaction
between nodes $i$ and $j$.
Additionally, in this model a network ``hub'' having  many connections is 
modeled as having a large value of $u_{i,1}^2+u_{i,2}^2$      
and makes most of its  connections to proteins having factors 
of smaller magnitude but in the same direction. 

The second panel of Figure \ref{fig:4} displays a different aspect of 
the protein network.   The plot identifies two groups of proteins
having large positive and large negative values of $\{u_{i,3}\}$
respectively. Members of each group are similar in the sense 
that they primarily interact with members of the opposite group but 
not with each other. 
The model captures this pattern
with a negative eigenvalue $\lambda_3$, so that $u_{i,3},u_{j,3}$ 
being large and of opposite sign is associated with a high probability
of interaction  between $i$ and $j$. 
In this way, the latent eigenmodel  
is able to  represent subnetworks that resemble bipartite graphs. 

For a detailed biological interpretation of the different hubs 
and clusters of the network, the reader is referred to 
\cite{butland_2005}.

\section{Discussion}
Distributions over the Stiefel manifold play an important role in 
spatial statistics, multivariate analysis and random matrix theory. 
An important family of probability distributions over this manifold is the 
matrix Bingham-von Mises-Fisher family, which generalizes the 
vector- and matrix-valued von Mises-Fisher  and Bingham distributions. 
This article has developed a rejection sampling scheme for the matrix
MF distribution and a Gibbs sampling scheme for the
matrix BMF distribution, thereby providing a
useful tool for  studying these complicated
multivariate probability distributions. 

Additionally, it has been shown 
that members of the BMF family of distributions arise as conditional 
posterior distributions in Gaussian and probit models for multivariate 
data. Likelihood-based approaches to multivariate modeling
 may be necessary  when the data are
 ordinal, missing or otherwise non-standard, and being able to sample 
from the BMF family allows for parameter estimation in these 
situations.



{\sf R}-functions to implement the sampling schemes outlined in 
this article are available at my website:
 \href{http://www.stat.washington.edu/hoff/}{\tt http://www.stat.washington.edu/hoff/}.

\bibliographystyle{plainnat}
\bibliography{hoff_bmv}

\end{document}